\documentstyle{numcool}
\begin{document}

\def\dim#1{\mbox{\,#1}}
\def\figdir{.}

\title[Incorporating Radiative Cooling]{Incorporating Radiative 
Cooling into a Cosmological Hydrodynamic Code}
\author[Gnedin]{Nickolay Y.\ Gnedin\\
University of California, Berkeley Astronomy Department,
Berkeley, CA 94720\\
e-mail: \sl gnedin@astron.berkeley.edu;
http://astron.berkeley.edu/~gnedin}

\def\capIE{
The gas density ({\it lower panel\/}), velocity ({\it middle panel\/}), and 
temperature ({\it upper panel\/}) for the exact spherically symmetric collapse
as a function of radius. Two moments are shown: the initial moment ({\it thin
line\/}) and the final moment ({\it bold line\/}) defined as the moment when 
the central density has increased by 12 orders of magnitude. The density
profile is a perfect power-law with the index -2 (except near the center where
numerical loss of accuracy occurs at very high densities).
}

\def\capIS{
The gas density ({\it lower panel\/}), velocity ({\it middle panel\/}), and 
temperature ({\it upper panel\/}) as a function of radius for the 
exact  case ({\it solid line\/}) and the softened gravity case ({\it
dashed line\/}). The softened case is shown for the time instant
$t=1.1t_{\rm col}$, where $t_{\rm col}$ is the collapse time of the
exact solution.
}

\def\capIC{
The gas cooling time ({\it lower panel\/}) and temperature ({\it upper
panel\/}) for three cases: exact solution ({\it solid line\/}),
softened gravity without cooling consistency condition ({\it dashed
line\/}), and softened gravity with the cooling consistency condition
({\it dotted line\/}). The balance between heating and cooling is
broken in the softened gravity solution which is manifested in the
unphysically long cooling times; the balance is restored back
in the consistent solution.
}

\def\capTT{
The gas cooling time vs dynamical time (in arbitrary units) for realistic
three-dimensional cosmological simulations of a standard CDM model: 
a 160 dynamical range simulation with 5\% baryons and without the 
cooling consistency condition ({\it upper left panel\/}); 
a 160 dynamical range simulation with 50\% baryons and without the 
cooling consistency condition ({\it lower left panel\/}); 
a 160 dynamical range simulation with 5\% baryons and with the 
cooling consistency condition ({\it upper right panel\/}); 
a 640 dynamical range simulation with 5\% baryons and without the 
cooling consistency condition ({\it lower right panel\/}); 
Again, the balance between heating and cooling is
broken without the consistency condition independently of the baryonic 
fraction; the balance is restored by the cooling consistency condition
or by increasing the resolution of a simulation. In the latter case however,
cooling and heating rates get out of balance again at the (finer) resolution 
limit of a higher resolution simulation.
}

\maketitle

\begin{abstract}
A possible inconsistency arising when a radiative cooling term is incorporated
in a finite resolution
self-gravitating hydrodynamic code is discussed. The inconsistency
appears when the heating-cooling balance within the cooling and collapsing 
gas cloud is broken near the resolution limit
of a numerical code. As the result, the cooling time of a fluid element
increases enormously, leading to the unphysical conclusion that the
fluid element does not cool and is therefore stable against the collapse.
A special {\it cooling consistency condition\/} is introduced which
approximately restores the heating-cooling balance and leads to a 
numerical solution that closely mimics the exact (infinite resolution)
solution.
\end{abstract}


%--------------------------------------------------------
%
%   Paper starts here
%
%--------------------------------------------------------

\section{Introduction}

Cosmological hydrodynamic simulations are now playing larger and larger
role in developing the theory of galaxy formation. They are successfully
incorporating more and more complex physical processes that are important
in the process of galaxy formation, such as dark matter evolution,
gas dynamics, radiative processes, star formation etc. Simulations, 
therefore, are challenged to solve a very complex system of nonlinear
equations incorporating physical processes with widely different 
characteristic time-scales. However, one usually does not solve the whole
system of equations by a single numerical technique (simply because there are
no such methods invented yet), but by combining various numerical techniques 
for separately solving for dark matter, gas dynamics, radiative cooling etc.
The advantage of this approach is that one can use known numerical methods
that have been developed to separately simulate each of physical processes
incorporated into a larger simulation. However, the pay off for using
different methods for solving parts of the whole problem is that those
methods are not necessarily consistent with each other; in other words,
straightforward combining of two of numerical techniques that solve
two different sub-sets of the whole system of nonlinear equations may
not lead to a techniques that solves the whole system. An example of how
direct merging of a gas dynamical solver with a gravity solver
can lead to a code that produces unphysical results (energy is not
balanced locally) is given in Gnedin \& Bertschinger \shortcite{GB96}.
Therefore,
a special {\it gravitational consistency condition\/} between the gravity and
gas dynamics ought to be satisfied in order to produce physically
sensible solutions.

However, the requirement of consistency between numerical methods
solving separate parts of the whole system of equations is a general
one, and applies not only to the case of gravity and gas dynamical
solvers. In particular, in this paper I demonstrate how a
{\it cooling consistency condition\/} can emerge when radiative
cooling is incorporated into a self-gravitating hydrodynamic code.

Radiative cooling is the crucial mechanism responsible for condensation
and collapse of baryonic gas into galaxies, and an adequate treatment of
it in numerical simulations is required in any realistic cosmological
gas dynamical simulation except, perhaps, in a simulation of clusters of
galaxies. Yet, since
radiative cooling can often introduce into the solution the time-scales that
are orders of magnitude shorter than the characteristic time-scale
for the gas evolution (the sound crossing time), no numerical
scheme has been invented to solve the whole system of equations including
gas dynamics and cooling together, but rather two different numerical
methods, one for gas dynamics, and another for cooling, are combined
together to incorporate radiative cooling into the gas dynamical code
\cite{C92,KHW92,ESD94,ANC94,G95}. Therefore, 
one can expect that consistency between those two different methods
may not be automatically achieved.

The cooling inconsistency was largely ignored in the previous numerical
work except in a simulation by Katz et al.\ \shortcite{KHW92}, who have 
abandoned the cooling time condition in their three-dimensional simulation of
galaxy formation for all gas elements below $30,000\dim{K}$; 
this solution to the cooling inconsistency
is acceptable as long as an equilibrium standard cooling curve is used
as a cooling function; in a simulation which includes nonequilibrium 
time-dependent ionization, radiative transfer effects, cooling by heavy 
elements and/or by molecules, this
approach would become inappropriate, since a cooling function is then a 
function of position and time, and no characteristic temperature like
$30,000\dim{K}$, below which the cooling time condition may be abandoned,
can be specified a priori.

This papers is composed as follows. In \S 2 I utilize the spherically
symmetric Lagrangian hydrodynamic code to emphasize the importance
and role of the cooling consistency condition, and in \S 3 I apply
those results to correct the existing cosmological hydrodynamic
code based on the ``Softened Lagrangian Hydrodynamics'' method 
(SLH; Gnedin 1995) for the inconsistency between the
hydrodynamic and cooling solvers. I conclude in \S4 with brief
discussion.

\section{Cooling Catastrophe in a Spherically Symmetric Case}

I first consider a collapse of a spherically symmetric selfgravitating
gas cloud
which is loosing its energy via radiative cooling. At the initial
state the cloud is at rest, has a uniform temperature $T=1$
(in dimensionless units) and the density profile:
\begin{equation}
	\rho = {1\over 1 + r^2},
	\label{rhoofr}
\end{equation}
where all quantities are dimensionless for simplicity, and the gravitational
constant $G=1$. The gas has
a constant polytropic index $\gamma=5/3$, and equation of state is
\[
	P = \rho T.
\]
The internal energy of the gas is lost due to radiative cooling at a rate
\begin{equation}
	\left(d U\over dt\right)_{\rm COOL} = -\rho^2\Lambda(T),
	\label{coolterm}
\end{equation}
where the cooling function $\Lambda(T)$ has the following form:
\begin{equation}
	\Lambda(T) \equiv 10^{-2} T^5\exp(-T^{-4}).
	\label{cf}
\end{equation}
This cooling function allows for an analytical solution for the
cooling evolution at a time interval short compared to the dynamical
time and simultaneously mimics the steep cut-off at $10^4\dim{K}$
in the real cooling function. 

Apparently, the gas cloud would loose its energy due to radiative cooling
and would collapse to the central singularity in a finite time, the process
known as the ``cooling catastrophe''. In order to simulate this process, the 
evolution 
of the gas is followed numerically by a one-dimensional Lagrangian
code as described in Gnedin \shortcite{G95} (adapted for the spherically
symmetric case and combined with the exact gravity solver). Since the 
hydrodynamic solver is exactly Lagrangian, the gravitational consistency
condition becomes trivial \cite{GB96}.

At each hydrodynamic time step $t_n$ new values of the gas density
$\rho_n$, velocity $v_n$, and temperature without cooling $\tilde T_n$ 
are computed
using the values at the previous time step $t_{n-1}$. 
Then, in order to account for 
radiative cooling, the following equation for the temperature is solved:
\begin{equation}
	{{\rm d}T\over{\rm d}t} = -(\gamma-1)\rho_n\Lambda(T)
	\label{coolsol}
\end{equation}
in a time interval $\Delta t=t_n-t_{n-1}$ with the initial condition 
$T(t_{n-1})=\tilde T_n$.\footnote{
With the cooling function (\ref{cf}) this equation
can be solved analytically.} 
This is the way the radiative cooling is incorporated
in most of existing cosmological hydrodynamic codes.

\begin{figure}
\par\centerline{%
\epsfxsize=1.0\columnwidth\epsfbox{\figdir/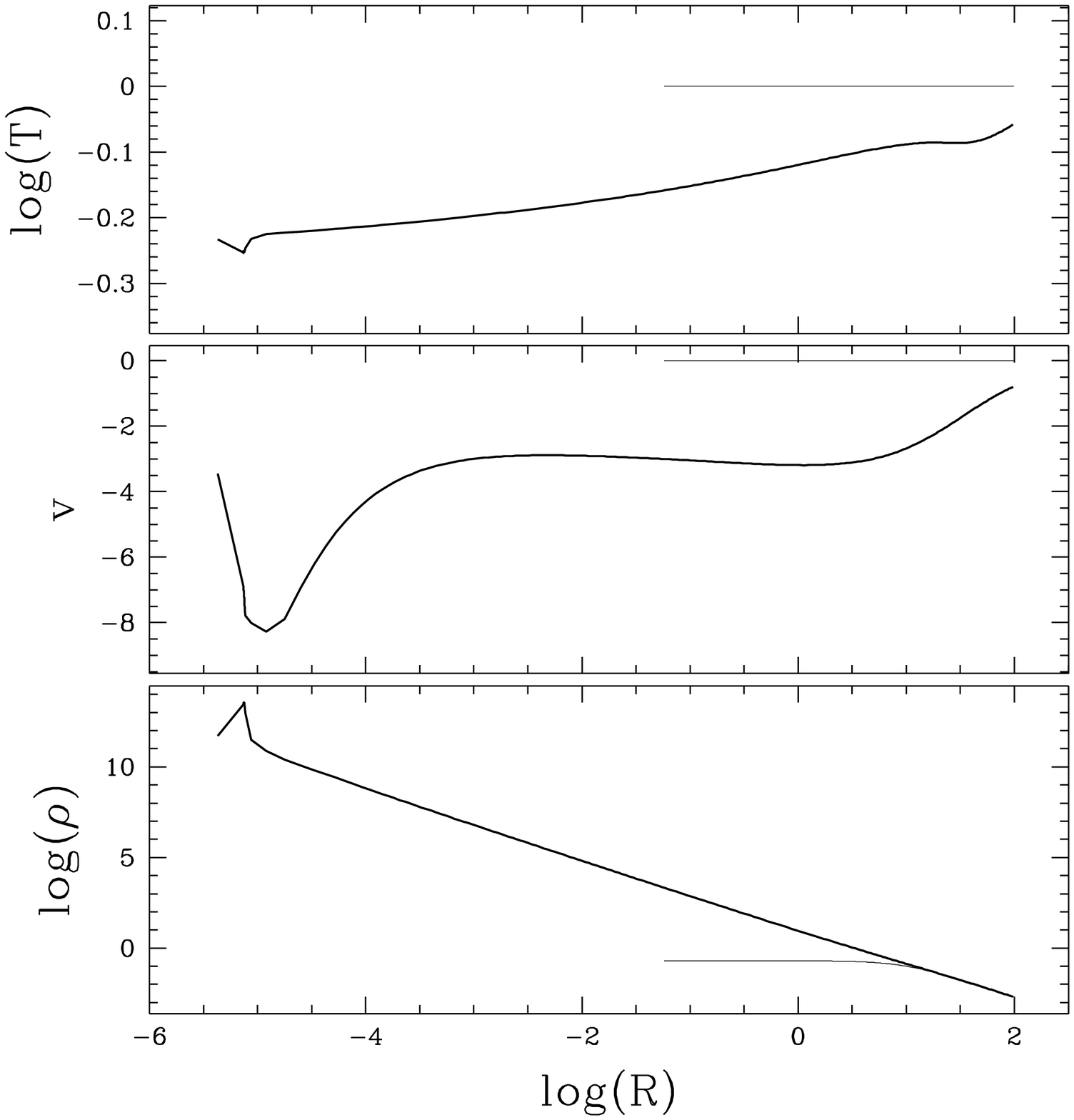}}%
\caption{\label{figIE}\capIE}
\end{figure}
Figure \ref{figIE} shows the gas density, velocity and
temperature as a function of radius at two different moments: the initial
moment $t=0$, shown as a thin solid line, and the moment when the central
density reached $10^{12}$, i.e.\ increased by 12 orders of the magnitude
(the bold line).
At this moment the simulation with 256 zones encountered numerical
problems at the center due to the round-off error in double
precision calculations. The density distribution is a perfect power law
with the index -2 across the  7 orders of magnitude in radius and 14
orders of magnitude in density, and the gas temperature is almost constant,
changing by less than $0.2\dim{dex}$ (60\%) across 7 orders of magnitude
in radius. 

This results demonstrates both the success and failure of the purely
Lagrangian gas dynamics. It definitely succeeds in achieving enormous
dynamical range, and simultaneously it fails to carry on the solution 
beyond the collapse time. Indeed this is a correct result which tells
one that the simple model one adopted is not sufficient to describe the
behavior of a real physical system. Nevertheless, in a real three
dimensional simulation, one would like to proceed further in time 
simulating other regions of the universe even if
there has formed a real singularity in one point in space. Numerically
this is achieved by softening the gravity solver, i.e.\ using a softened
gravity law instead of the exact $1/r^2$ Newtonian law. 

\begin{figure}
\par\centerline{%
\epsfxsize=1.0\columnwidth\epsfbox{\figdir/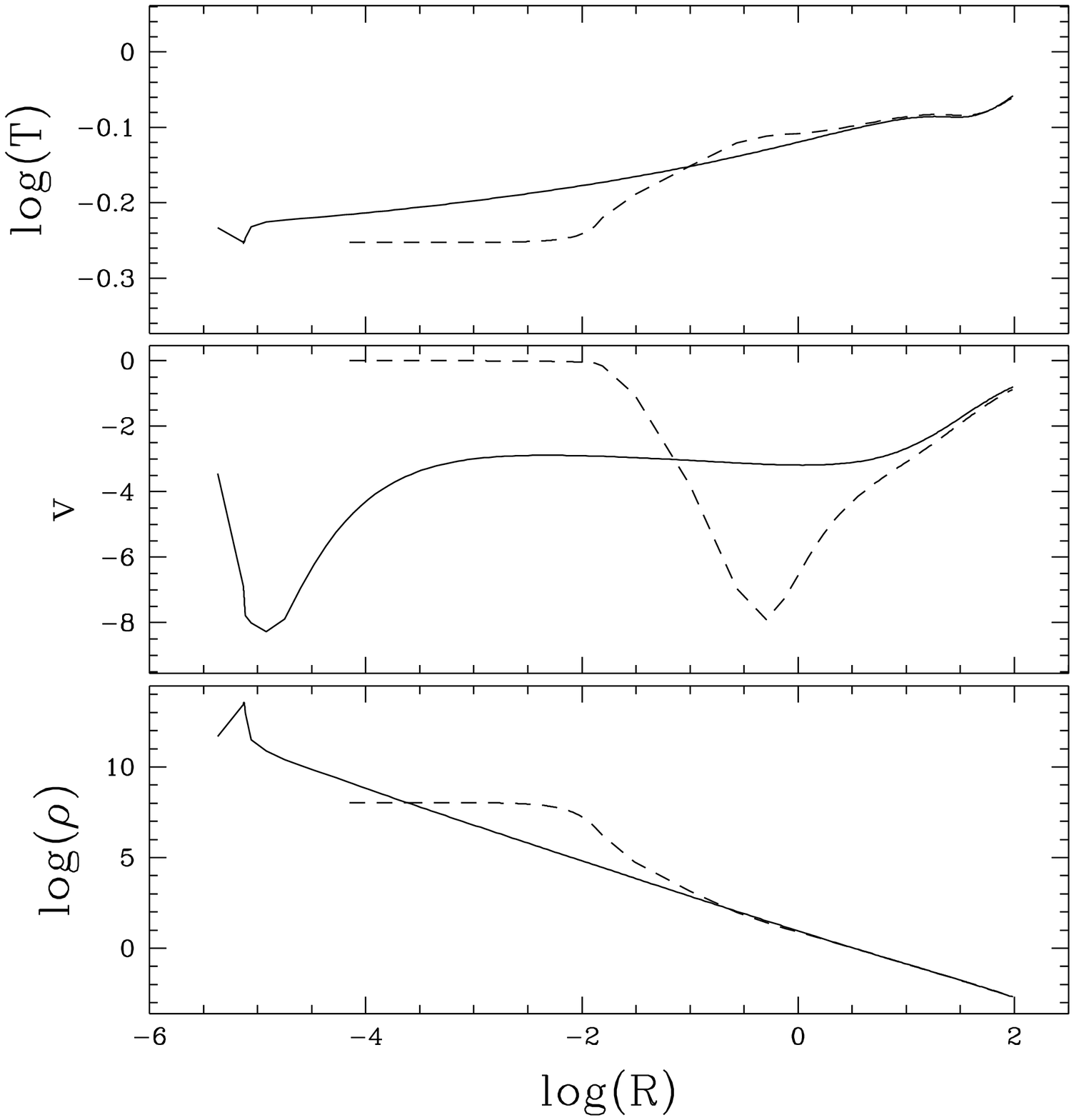}}%
\caption{\label{figIS}\capIS}
\end{figure}
Let me now consider the cooling catastrophe for the softened gravity case.
The standard Plummer softened gravity law is adopted with the softening
length $\epsilon=0.01$ and the same initial
configuration is followed forward in time. In this case no
real singularity forms, and the solution can be carried on in time
indefinitely.
Figure \ref{figIS} shows the gas density, velocity and the temperature
for the exact case as in Fig.\ref{figIE} with a solid line, and for
the softened case with the dashed line. The softened solution is shown
at a moment $t=1.1t_{\rm col}$, where $t_{\rm col}$ is the time of collapse
for the exact solution. As one can expect, the velocity approaches zero,
and the density and the temperature approach finite values
at the center, with the core radius close to the softening length $\epsilon$.

As one can expect, the softened solution provides a good approximation to the
real solution on scales larger than several softening lengths. The temperature 
inside the softening length is only slightly lower than the exact solution
and thus we may conclude that the softened solution is a good approximation
to the exact solution. Here, however, the cooling consistency condition
steps in.

When the gravitational consistency condition is violated, there is a simple
basic law of physics - energy conservation - which is also violated, and it 
is then clear that the gravitational consistency condition has to assure
the conservation of energy. It is not quite so with the cooling consistency
condition, since there is no basic physical law that is violated by the
softened solution in Fig.\ref{figIS}. If, for example, one is satisfied with
the mere existence of the gas globs described by the dashed line in
Fig.\ref{figIS} in one's simulation, one can simply ignore any cooling
inconsistency that may appear in the softened solution.
However, if one does not restrict himself to the mere existence of those
gas globs, but
is interested whether those objects actually cool and collapse to eventually
form stars (as the exact solution does), the softened solution becomes
unacceptable. 
\begin{figure}
\par\centerline{%
\epsfxsize=1.0\columnwidth\epsfbox{\figdir/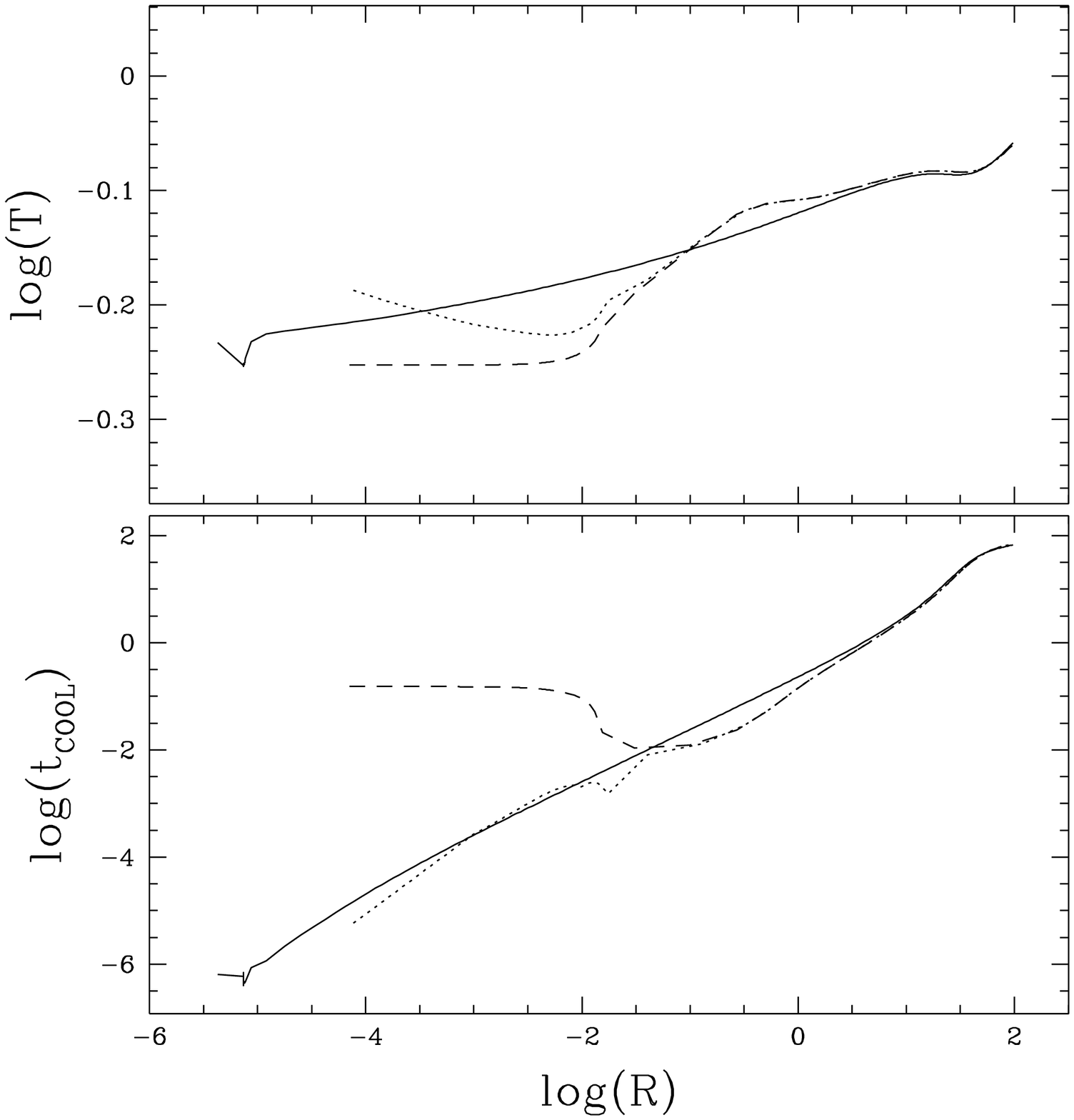}}%
\caption{\label{figIC}\capIC}
\end{figure}
In Figure \ref{figIC} I show the cooling time and the
temperature as a function of radius for the exact solution (the solid line)
and the softened solution (the dashed line; the dotted line is explained
below). Even if temperatures of two solutions differ by a small amount,
the cooling function is so steep, that the cooling time of the softened
solution exceed the cooling time of the exact solution by many orders of
magnitude. At every point in the exact solution the cooling time
$t_{\rm COOL}$ is similar to the local dynamical time $t_{\rm DYN}$,
\[
	t_{\rm COOL} \sim t_{\rm DYN}.
\]
This condition is satisfied because the collapse proceeds at the
dynamical time, and if the cooling time is much shorter than
the dynamical time, the temperature
will quickly decrease, the cooling function will also decrease and the
cooling time will increase until it is close to the dynamical time; if,
inversely, the dynamical time is shorter than the cooling time, the 
temperature will increase adiabatically, the cooling function will increase
and the cooling time will decrease until it is again approximately equal
to the dynamical time. The collapse is therefore proceeding in the
quasi-equilibrium state, when adiabatic heating is approximately
balanced by radiative cooling. 

The softened solution, however, grossly violates this quasi-equilibrium.
Since the softened gravity prevents the collapse much beyond the resolution
limit of the code, there is no adiabatic heating to balance  cooling, the
gas temperature will continue to decrease and the cooling time will
continue to increase as long as the simulation progresses, so that at
a sufficient late time moment $t$,
\[
	t_{\rm COOL} \sim t \gg t_{\rm DYN}.
\]
Then, if one tries to estimate whether the object is actually cooling
and collapsing by comparing its dynamical time to its cooling time, one
would find that the cooling time is several orders of magnitude larger
than the dynamical time, and would erroneously conclude that the object
does not cool and, therefore, does not collapse!

This conclusion stems from the fact that the hydrodynamic and cooling
solvers are not consistent with each other in a sense that the cooling
solver (\ref{coolsol}) knows nothing about whether the resolution limit
of the code has been already reached or not. I can therefore 
immediately conclude that for a fully Lagrangian solver the
cooling consistency condition is trivial as is the gravitational
consistency condition. However, in this case it is assumed that the
fully Lagrangian solver obeys the $1/r^2$ gravity law, whereas for
the gravitational consistency condition only potentiality of the
force is required \cite{GB96}.

Let me now consider the softened case and address the question whether
we can repair the deficiency of the cooling solver and make it
consistent with the (self-gravitating) hydrodynamic solver. Since the
inconsistency appears as the ``overcooling'' problem when the resolution
limit is reached, the obvious solution is to reduce the cooling rate
at the resolution limit of the code. I therefore replace the radiative 
cooling term (\ref{coolterm}) with the following expression:
\begin{equation}
	\left({\rm d}U\over{\rm d}t\right)_{\rm COOL} = -D\rho^2\Lambda(T),
	\label{modcoolterm}
\end{equation}
where $D$ is the {\it cooling consistency factor\/}, $0<D<1$.
Obviously, $D$ should be a function of the softening length $\epsilon$, 
so that $D=1$ when $r\gg\epsilon$ and $D=0$ if $r\ll\epsilon$.
To demonstrate the effect of the cooling consistency condition, I adopt
the following form for the cooling consistency factor $D$:
\begin{equation}
	D(r,\epsilon) = \left(r^2\over r^2+a\epsilon^2\right)^b,
	\label{ddef}
\end{equation}
where
\begin{equation}
	a=\cases{1, & if $r<0.1\epsilon$, \cr
	0.01, & if $r>\epsilon$, and \cr
	0.01(\epsilon/r)^2, & otherwise,}
	\label{ddefa}
\end{equation}
and
\begin{equation}
	b=\cases{0.75, & if $r<0.01\epsilon$, \cr
	0.85, & if $r>0.1\epsilon$, and \cr
	0.75+0.1(2+\log_{10}(r/\epsilon)), & otherwise.}
	\label{ddefb}
\end{equation}
The dotted line in Fig.\ref{figIC} shows the temperature and the
cooling time for the softened case when the cooling consistency
condition is incorporated. The cooling time agrees very well with
the exact solution. The density and the velocity profiles for
the corrected case coincide with the uncorrected softened case
almost precisely, so the cooling consistency condition does not
change the density profile. Apparently, it is possible to achieve even 
better agreement by an appropriate choice of the cooling consistency
factor $D$. The particular choice presented above serves merely to
demonstrate the importance and effect of the cooling consistency
condition.

\section{Cooling Consistency Condition in Three Dimensions}

Let me first briefly discuss the rationale behind introducing the
cooling consistency factor $D$. The equation for the temperature
of a fluid element with a constant polytropic index $\gamma$
can be written as:
\begin{equation}
	{{\rm d}T\over{\rm d}t} = (\gamma-1)\left[{T\over\rho}
	{{\rm d}\rho\over{\rm d}t} - \rho\Lambda(T,\rho)\right],
	\label{lageq}
\end{equation}
where $d/dt$ denotes the Lagrangian time-derivative, and the cooling
function $\Lambda$ is not required to be a function of $T$ only.
For a cooling catastrophe, the dynamical time is close to the cooling
time, which implies that the two terms in square brackets almost cancel
each other and the collapse is almost isothermal:
\begin{equation}
	{T\over\rho}
	{{\rm d}\rho\over{\rm d}t} \approx \rho\Lambda(T,\rho).
	\label{lagbal}
\end{equation}
When equation (\ref{lageq}) is implemented numerically, the time
derivative becomes numerical time derivative for a finite resolution
numerical scheme, which I emphasize by adding a subscript $N$ to it:
\begin{equation}
	{{\rm d}_NT\over{\rm d}t} = (\gamma-1)\left[{T\over\rho}
	{{\rm d}_N\rho\over{\rm d}t} - \rho\Lambda(T,\rho)\right].
	\label{numeq}
\end{equation}
The main difference between equations (\ref{lageq}) and (\ref{numeq}) is
that at the resolution limit of the code the density stops changing,
\begin{equation}
	{{\rm d}_N\rho\over{\rm d}t}(\mbox{res.\ limit}) \rightarrow 0,
\end{equation}
and equation (\ref{numeq}) becomes:
\begin{equation}
	{{\rm d}_NT\over{\rm d}t} = -(\gamma-1)
	\rho\Lambda(T,\rho),
	\label{reseq}
\end{equation}
implying that the cooling time will
increase in proportion to the physical time $t$ and not the dynamical
time $t_{\rm DYN}$. The equation (\ref{numeq}) can be ``repaired''
by introducing the cooling consistency factor $D$,
\begin{equation}
	{{\rm d}_NT\over{\rm d}t} = (\gamma-1)\left[{T\over\rho}
	{{\rm d}_N\rho\over{\rm d}t} - D\rho\Lambda(T,\rho)\right].
	\label{coneq}
\end{equation}
Now, by requiring that
\begin{equation}
	D = {{\rm d}_N\rho/{\rm d}t\over{\rm d}\rho/{\rm d}t},
	\label{ccc}	
\end{equation}
I restore the quasi-equilibrium character of the collapse (eq. 
[\ref{lagbal}]). The only disadvantage of this approach is that the
exact cooling consistency condition (\ref{ccc}) can never be
satisfied since the exact time derivate of the density,
$d\rho/dt$ is unknown, and more that that, it is what one tries to
approximate with the numerical derivative ${\rm d}_N\rho/{\rm d}t$. 
The cooling
consistency condition can therefore be satisfied only {\it approximately\/}.

Reducing cooling is not the only way to achieve cooling consistency.
It is possible, for example, to introduce additional heating terms in
equation (\ref{numeq}) to account for the loss of adiabatic heating near
the resolution limit of the code. Therefore, there exist many
different ways to satisfy the cooling consistency condition, and the
cooling consistency factor $D$ is only one of them.

\begin{figure}
\par\centerline{%
\epsfxsize=1.2\columnwidth\epsfbox{\figdir/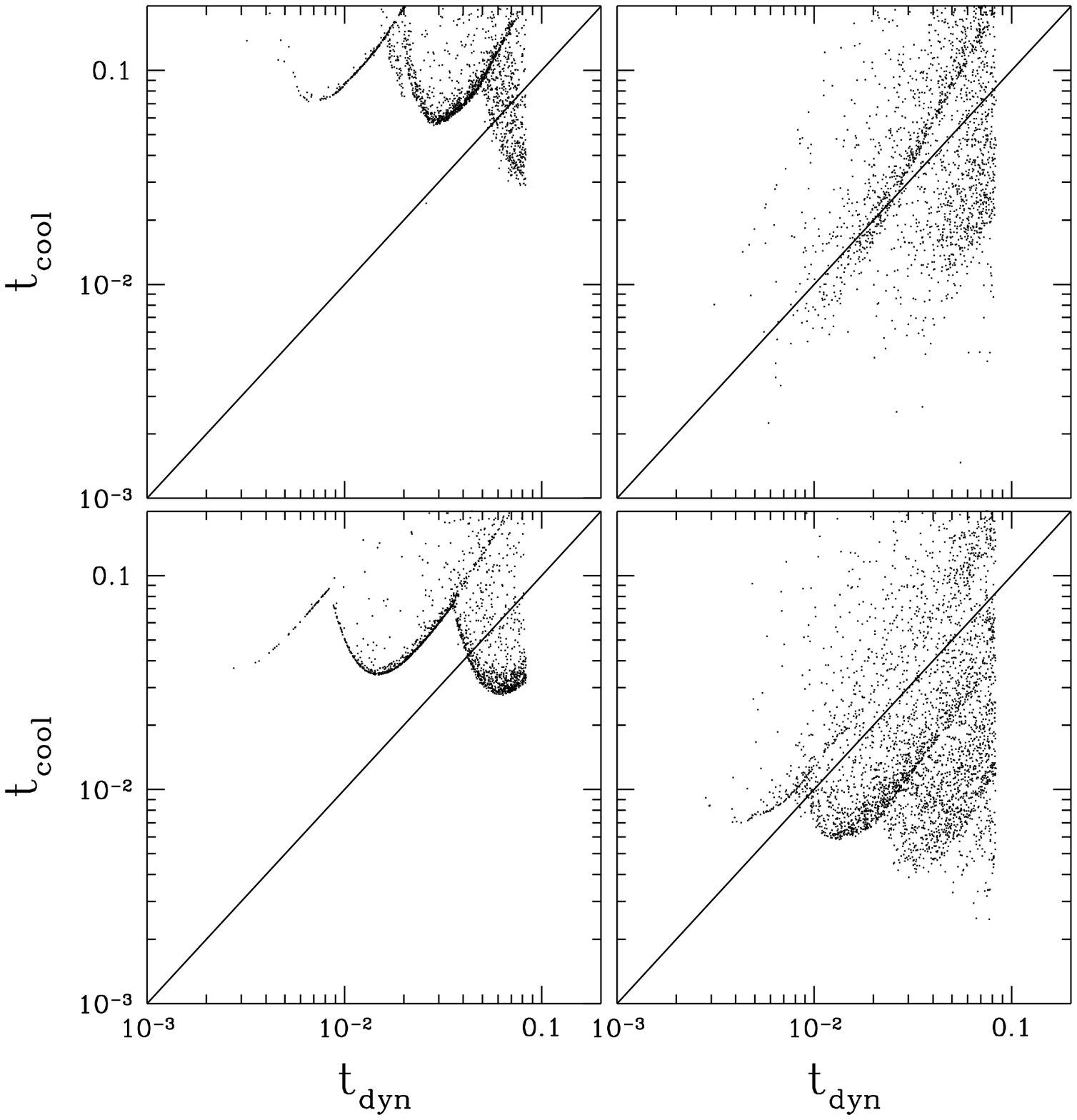}}%
\caption{\label{figTT}\capTT}
\end{figure}
Let me now turn to a three-dimensional case. I use the
SLH-P$^3$M code \cite{GB96} to
demonstrate the effect of the cooling consistency condition on the 
thermal evolution of the cosmic gas. I
adopt the standard CDM model as a testbed for my demonstration. The 
cosmological parameters are fixed so that $\Omega_0=1.0$, $h=0.5$, and
I fix the initial power spectrum by using the BBKS transfer function
\cite{BBKS86} and adopting $\sigma_8=1.0$. 

First, two $32^3$
runs with the softening parameter of $1/5$ (the dynamical range of 160)
and {\it without\/} any consistency 
condition were performed, the first one
using $\Omega_b=0.05$, and, therefore, having 5\% baryons, and the second
one using $\Omega_b=0.5$ (50\% baryons) with otherwise identical initial
conditions. Figure \ref{figTT} shows cooling
and dynamical times for all fluid elements with overdensities in excess of
100 and temperatures between $10^{3.5}\dim{K}$ and $10^{4.5}\dim{K}$
at $z=6$ in a $2h^{-1}\dim{Mpc}$ box (at this moment most of the gas has not 
yet been shocked to temperatures significantly exceeding $10^4\dim{K}$)
for those two runs in the upper left and lower left panels respectively.
It is apparent that the cooling time in the highest density regions is very
long, exceeding the dynamical time by up to two orders of magnitude, 
and the balance between the cooling and heating rates is severely broken -
the manifestation of the cooling inconsistency.
Were those fluid elements tested
on the condition of collapse, one would find that they cannot cool, and,
therefore, cannot collapse, which is apparently nonphysical conclusion.
As one can expect, the cooling inconsistency does not depend on the
baryonic fraction ((and, hence, on the dark matter fraction) since
equation (\ref{numeq}) contains no reference to the dark matter fraction.

I can now introduce the {\it SLH cooling consistency condition\/} by defining
the cooling consistency factor $D$ in the following way:
\begin{equation}
	D = 1-\mbox{min}(\sigma_1,\sigma_2,\sigma_3),
	\label{slhccc}
\end{equation}
where $\sigma_j$ are eigenvalues of the deformation tensor
$\sigma^{ij}$ which defines the resolution limit of the SLH code
(for exact definitions see Gnedin \& Bertschinger \shortcite{GB96}).
The cooling consistency condition in this form gradually
reduces the total cooling rate as the gas element followed by the SLH code
approaches the code resolution limit. 

Since the SLH code is not exactly Lagrangian, and, in 
general, the
flow is not spherically symmetric, an expression
for $D$ found in the spherically symmetric
case (eq. [\ref{ddef}]) cannot be applied in a three-dimensional case exactly.
However, in the spherically symmetric case, the expression (\ref{slhccc}) 
reduces to equation (\ref{ddef}) with $a=b=1$. While corrections 
(\ref{ddefa}) and (\ref{ddefb}) improve the accuracy of the approximate
solution, they cannot be easily generalized for a three-dimensional case.
The three-dimensional consistency condition (\ref{slhccc}) is, therefore,
less accurate that the three-dimensional approximation given by (\ref{ddef});
further improving upon the accuracy of the SLH consistency condition is beyond
the frame of this paper.

I can now test the SLH consistency condition (\ref{slhccc}).
The distribution of cooling and
dynamical times for a $32^3$ simulation with the $1/5$ softening and 
$\Omega_b=0.05$ {\it with\/} the cooling consistency condition as given by
equation (\ref{slhccc}) is shown in Fig.\ref{figTT} at the upper right panel.
One can note now that in the run with the consistency
condition, the cooling times are in a reasonable
(yet not perfect, since the cooling consistency condition (\ref{slhccc})
is only an approximate one) agreement with the dynamical times for
all fluid elements with high density. For those fluid elements the cooling
time is of the order of their dynamical time and they are {\it apparently\/} 
cooling and collapsing. 
One must remember, of course, that in the simulation
most of those fluid elements have already reached their resolution limit
and their density does not significantly increase with time, but the
relationship between their cooling and dynamical times mimics that of
truly cooling and collapsing gas clouds.
A simulation with the cooling consistency condition
and $\Omega_b=0.5$ (50\% baryons) produces nearly identical distribution
of cooling and dynamical times and is not shown here due to the space
limitations.

Finally, the lower right panel of Fig.\ref{figTT} shows the distribution of
cooling and dynamical times for a $64^3$ simulation with the softening 
parameter of $1/10$ (the dynamical range of 640) with the same $2h^{-1}
\dim{Mpc}$ box and $\Omega_b=0.05$ {\it without\/} the cooling consistency
condition (for clarity only every eights fluid element is shown). 
Since this simulation has four times higher spatial resolution
than the simulation shown in the upper left panel of Fig.\ref{figTT},
the balance between cooling and heating extends to about a factor
of 64 higher densities, or, equivalently, to about a factor of 8 smaller
dynamical times. One can easily see that the simulation with the consistency 
condition (the upper right panel) mimics very closely the high resolution
simulation.
At the highest densities the balance is again broken,
as manifested by the cooling time becoming larger than the dynamical
time as the dynamical time decreases. The effect is not as dramatic
as in the upper left panel, and the gas densities in the high resolution 
simulation are only slightly larger than in the low resolution simulations
because the final epoch ($z=6$) was chosen such that in the high 
resolution simulation
only a small number of all fluid elements have actually reached their
resolution limit; this makes the high resolution simulation closely 
mimicking an imaginary exact (infinite resolution) case.

\section{Conclusions}

I have shown on simple spherically symmetric and realistic three-dimensional
examples how the inconsistency between the finite resolution self-gravitating
hydrodynamic solver and the radiative cooling term may arise. For a
efficiently cooling and collapsing gas clouds the cooling time approximately
balances the dynamical time so that the collapse is occurring in the
quasi-equilibrium between adiabatic heating and radiative cooling. 

In simulations, the finite resolution leads to decrease in adiabatic heating
when a fluid elements approaches the resolution limit of a numerical
code, and the balance between heating and coolings breaks down. This leads
to an ``overcooling'' problem, when the cooling time in the fluid element
becomes significantly longer than the dynamical time, and the fluid element
appears cooling very inefficiently. 

The contradiction is eliminated when the cooling consistency condition
is introduced, which reduces the cooling rate in proportion to the
reduction in the heating rate. I present the exact cooling consistency
condition, which however cannot be realized in practice since it depends
on the unknown true solution which a simulations tries to approximate.
However, approximate cooling consistency conditions can be introduced
both in a spherically symmetric and realistic three-dimensional cases
which are ``good enough'' in a sense that the resultant approximate
numerical solution mimics the exact solution in balancing the
heating and cooling rates.

This work was supported in part by the 
NSF grant AST-9318185
awarded to the Grand Challenge Cosmology Consortium and the
UC Berkeley grant 1-443839-07427.
Some of the simulations was performed on the NCSA Power Challenge
Array under the grant AST-960015N.

\end{document}